\documentclass{aa}
\usepackage[varg]{txfonts}

\usepackage{natbib}
\bibpunct{(}{)}{;}{a}{}{,} 

\usepackage{graphicx}
\usepackage{txfonts}
\newcommand{\comment}[1]{}
\newcommand{\vect}[1]{\mathbf{#1}}
\newcommand{\changed}[1]{#1}

\begin{document}
 
\title{The Single-mode Complex Amplitude Refinement (SCAR) coronagraph}
\subtitle{I. Concept, theory and design}
\author{E.H.~Por \and S.Y.~Haffert}
\institute{Sterrewacht Leiden, PO Box 9513, Niels Bohrweg 2, 2300 RA Leiden, The Netherlands\\
\email{por@strw.leidenuniv.nl}}
\date{Received: <date>; accepted: <date>}
\abstract
{The recent discovery of an Earth-mass exoplanet around the nearby star Proxima Centauri provides a prime target for the search for life on planets outside our solar system. Atmospheric characterization of these planets has been proposed by blocking the starlight with a stellar coronagraph and using a high-resolution spectrograph to search for reflected starlight off the planet.}
{Due to the large flux ratio and small angular separation between Proxima b and its host star ($\lesssim10^{-7}$ and $\lesssim2.2\lambda/D$ respectively; at 750nm \changed{for an 8m-class telescope}) the coronagraph needs to have a high starlight suppression at extremely-low inner-working angles. Additionally, it needs to operate over a broad spectral bandwidth and under residual telescope vibrations. This allows for efficient use of spectroscopic post-processing techniques. We aim to find the global optimum of an integrated coronagraphic integral-field spectrograph.}
{We present the Single-mode Complex Amplitude Refinement (SCAR) coronagraph that uses a microlens-fed single-mode fiber array in the focal plane downstream from a pupil-plane phase plate. The mode-filtering property of the single-mode fibers allows for the nulling of starlight on the fibers. The phase pattern \changed{in the pupil plane} is specifically designed to take advantage of this mode-filtering capability. Second-order nulling on the fibers expands the spectral bandwidth and decreases the tip-tilt sensitivity of the coronagraph.}
{The SCAR coronagraph has a low inner-working angle ($\sim 1\lambda/D$) at a contrast of $<3\times10^{-5}$ for the 6 fibers surrounding the star using a sufficiently-good adaptive optics system. It can operate over broad spectral bandwidths ($\sim 20\%$) and delivers high throughput ($>50\%$ including fiber injection losses). Additionally, it is robust against tip-tilt errors ($\sim0.1\lambda/D$ rms). \changed{We present SCAR designs for both an unobstructed and a VLT-like pupil.}}
{The SCAR coronagraph is a promising candidate for exoplanet detection and characterization around nearby stars using current high-resolution imaging instruments.}


\maketitle

\section{Introduction}
\label{sec:introduction}

The discovery of many rocky exoplanets around stars \changed{\citep{borucki2011characteristics}} has prompted the radial velocity search of the closest and brightest ones. This led to the discovery of a terrestrial exoplanet in the habitable zone around Proxima Centauri~\citep{anglada2016terrestrial}. This planet does not transit its host star~\citep{kipping2017no}, making transit spectroscopy impossible. Proxima b however has an angular separation in quadrature of $\sim2.2\lambda/D$ at 750nm \changed{for an 8m-class telescope}, making a spatially resolved imaging approach feasible. \cite{lovis2017atmospheric} explored the possibility of coupling the high-contrast imager SPHERE~\citep{beuzit2008sphere} to the high-resolution spectrograph ESPRESSO~\citep{pepe2010espresso}. The implementation of the coronagraph was left as an open question. Here we show a new type of coronagraph that will enable a combination of SPHERE and a high-resolution spectrograph to successfully observe Proxima b.

With the advent of extreme adaptive optics systems (xAO), such as SPHERE \citep{beuzit2008sphere}, GPI \citep{macintosh2014first} and SCExAO \citep{jovanovic2015subaru}, direct detection has taken some major steps forward. These systems create a diffraction-limited point spread function (PSF), which allows for the use of coronagraphy to enhance the contrast of observations. Coronagraphs need to suppress stellar light at the location of the planet, while having high throughput for the planet itself. Additionally, they need to operate over a broad wavelength range and in the presence of residual telescope aberrations, both static and dynamic. Telescope vibrations in particular turned out to be a major concern for these high contrast imaging instruments \citep{fusco2016saxo}. Pupil-plane-only coronagraphs are preferable, because their performance is independent of telescope pointing making telescope vibrations less of a concern. An inherent disadvantage \changed{of pupil-plane coronagraphs} is that the coronagraphic throughput loss is the same for the star and the planet. Therefore designs \changed{of pupil-plane coronagraphs} to reach high contrasts or extremely-low inner working angles intrinsically have a low throughput. As a result, coronagraphs combining focal-plane and pupil-plane optics often outperform pupil-plane-only coronagraphs at extremely-low inner-working angles \citep{mawet2012review}.

Even with the best adaptive optics systems, residual aberrations will always limit the raw contrast of ground-based observations to $\sim10^{-6}$ to $10^{-7}$ \citep{guyon2005limits}. Currently however, observations are limited by the non-common-path errors between the wavefront sensor and science camera, creating quasi-static speckles in the focal-plane. These speckles amplify atmospheric residuals \citep{aime2004usefulness} and are notoriously hard to predict. Post-processing of coronagraphic images can enhance the contrast beyond the raw-contrast limit. Diversity of some kind is often used to calibrate the instrument itself. Angular diversity \citep{marois2006angular} uses the rotation of the sky with respect to the instrument and has provided excellent results. A recent development in this field uses the diversity in radial velocities of the star and the planet: stellar-light speckles still retain the radial velocity of the star, while the planet appears at a different velocity altogether. \changed{Cross-correlation techniques on high-resolution spectroscopy \citep{sparks2002imaging,riaud2007improving,konopacky2013detection} combined with coronagraphy \citep{kawahara2014spectroscopic,snellen2015combining,wang2017observing} have shown great promise. They have provided dayside spectroscopy of $\tau$ Bo\"{o}tis b \citep{brogi2012signature} and even the first measurement of planetary spin \citep{snellen2014fast}.}

Integration of both coronagraphy and high-resolution spectroscopy into a single concept has only recently been attempted. \cite{mawet2017observing} used a fiber injection unit in the focal-plane downstream from a conventional vortex coronagraph. A single-mode fiber was centered around the planet and the stellar light speckles were removed using active speckle control algorithms. While this setup does allow for transportation of the light to a dedicated high-resolution spectrograph, it does not \changed{optimally} combine both methods. In this paper we take the next step: we allow the coronagraph design to depend on the single-mode fibers in the focal plane. This \changed{allows for more freedom in} the design \changed{process} and provides better coronagraphic throughput as those modes filtered out by the fiber injection unit \changed{do not} need to be suppressed by the \changed{upstream} coronagraph.

\changed{Interestingly, the use of single-mode fibers for coronagraphy is not new. \cite{haguenauer2006deep} already proposed using a single-mode fiber to null the star by using a $\pi$ phase shift on part of the pupil. This was further developed by \cite{martin2008development} and finally put on sky by \cite{hanot2011improving}. These applications of single-mode fibers for coronagraphy were based on interferometry. \cite{mawet2017observing} were the first to put single-mode fibers behind a conventional coronagraph.}

In Sect.~\ref{sec:nulling} we describe nulling on single-mode fibers, extend the fiber injection unit to use multiple single-mode fibers and show the coronagraphic capabilities and throughput of such a system. In Sect.~\ref{sec:coronagraphy} we use an apodizing phase plate coronagraph to expand the spectral bandwidth and decrease the tip-tilt sensitivity. In Sect.~\ref{sec:properties} we describe the throughput, inner-working angle, chromaticity and sensitivity to aberrations of this new system. We conclude with Sect.~\ref{sec:conclusion}.

\section{Modal filtering using single-mode fibers}
\label{sec:nulling}
\comment{Fiber introduction and MLA: what does a fiber do. How can we gain contrast. Explain method. Many fibers in an array to sample focal plane. Use MLA to separate fiber cores and increase throughput stability and continuous field of view. This is still mode-filtering in the focal plane. Show throughput with normal PSF. Show throughput with existing coronagraphs (specifically APP).}

\subsection{Nulling in single-mode fibers}

The coupling efficiency $\eta_\mathrm{sm}$ of light into a single-mode fiber can be calculated by the projection of the input electric field $E_\mathrm{in}$ onto the mode of the fiber $E_\mathrm{sm}$ as
\begin{equation}
\eta_\mathrm{sm} = \frac{\left| \int E^*_\mathrm{in} E_\mathrm{sm} \mathrm{d}A \right|^2}{\int |E_\mathrm{in}|^2 \mathrm{d}A~\int |E_\mathrm{sm}|^2 \mathrm{d}A},
\label{eq:fiber_coupling}
\end{equation}
where the integration is done over all space. The fiber mode $E_\mathrm{m}$ can be calculated using waveguide theory and the geometry of the fiber in question, but in this paper we use the Gaussian approximation \citep{marcuse1978gaussian}
\begin{equation}
E_\mathrm{sm}(r) = \exp{\left[-\frac{r^2}{w^2}\right]}
\end{equation}
where $r$ is the distance from the center, and $2w$ is the mode field diameter of the fiber. We see that the coupling efficiency $\eta_\mathrm{sm} \leq 1$ for all input fields and that maximum coupling is only attained when $E_\mathrm{in}$ matches the fiber mode.

\begin{figure}
\includegraphics[width=\columnwidth]{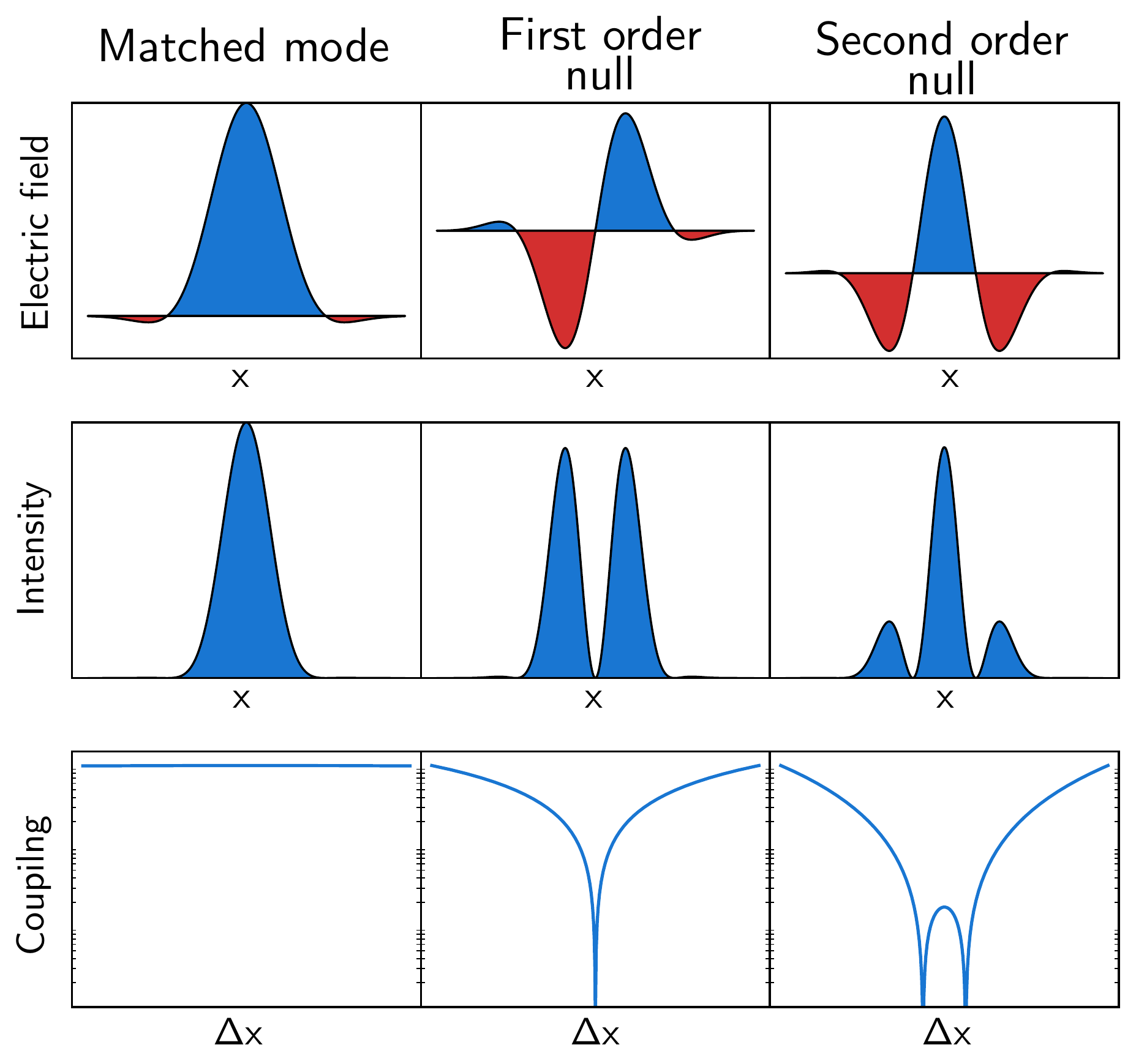}
\caption{Coupling into a single-mode fiber with a (a) matched electric field, (b) first order null and (c) second order null. The first row shows the electric field projected onto the fiber. The second row shows the intensity on the fiber face. The third row shows the coupling efficiency for off-center fibers. The matched mode couples well into the fiber, even for small off-center positions. The first-order null has no throughput: its odd electric field ensures a zero overlap integral in Eq.~\ref{eq:fiber_coupling}. Off-center positions however still transmit because the odd structure is lost. The second-order employs an even electric field where the contribution of the central peak is canceled by the two sidebands. This creates a much broader null when decentering the fiber.}
\label{fig:principle}
\end{figure}

Suppose now that we put a single-mode fiber in the focal plane of a telescope, with its mode field diameter matched to that of the Airy core of the PSF. \changed{Using Equation~\ref{eq:fiber_coupling} we can calculate the coupling efficiency $\eta_s(\vect{x})$ of the star as a function of focal-plane position $x$. We can do the same thing for the planet, yielding $\eta_p(\vect{x}, \vect{x}_0)$ where $\vect{x}_0$ is the location of the planet. The raw contrast at the fiber output can be written as
\begin{equation}
C_\mathrm{raw}(\vect{x}, \vect{x}_0) = \frac{\eta_\mathrm{p}(\vect{x}, \vect{x}_0) }{\eta_\mathrm{s}(\vect{x})}.
\end{equation}
Note that, when} the fiber is centered around the planet\changed{, ie. $\vect{x}=\vect{x}_0$}, the electric field of the planet will couple efficiently into the fiber, as the Airy core is closely matched to the Gaussian fiber mode. The electric field of the star at this position will however consist of Airy rings. These will be smaller, not only in intensity but also spatially so that \changed{around} two Airy rings will be visible on the fiber face. \changed{This is possible as the Airy core itself has a size of $\sim1\lambda/D$ full-width half-maximum (FWHM), while the Airy rings are sized $\sim0.5\lambda/D$ FWHM.} As neighboring Airy rings have opposite phase, the light from the two Airy rings will (partially) cancel each other in the projection integral of Eq.~\ref{eq:fiber_coupling}, resulting in a lower stellar throughput. This nulling provides an additional contrast enhancement not possible with multi-mode fibers. Fig.~\ref{fig:principle} illustrates this graphically in columns 1 and 2.

Column 3 in Fig.~\ref{fig:principle} shows an alternative null structure. This second-order null balances the contribution of the core with its two sidebands. This has the effect of broadening the null for decenter as the loss in overlap with one of the sidebands is compensated by the increase in overlap with the other. Note that in this case we have split the second-order null into two first-order null by subtracting a tiny fraction of the matched mode. This gives a characteristic double dip in the coupling curve and broadens the null even more by raising the coupling between the two first-order nulls. This coupling at the center must be kept below the design coupling.

\changed{This second-order null is the basis for the SCAR coronagraph. For comparison with existing coronagraph implementations, we define the ``integration time gain'' as the ratio between the integration time for unresolved imaging and coronagraphic imaging to reach a predefined signal-to-noise ratio. This can be expressed in the star and planet throughput as
\begin{equation}
\frac{\Delta T_\mathrm{coronagraphic}}{\Delta T_\mathrm{unresolved}} = \frac{\eta_s(\vect{x})}{\eta_p^2(\vect{x}, \vect{x}_0)}.
\end{equation}
This metric takes into account both the raw contrast and planet throughput of the coronagraph. Note that noise sources other than photon noise were ignored in this respect. As these only become important for small planet throughputs, we will show both the integration time gain $\eta_s/\eta_p^2$ and the planet throughput $\eta_p$.}

\subsection{Single-mode fiber arrays using microlenses}
\label{sec:single_mode_fiber_array}

\begin{figure}
\includegraphics[width=\columnwidth]{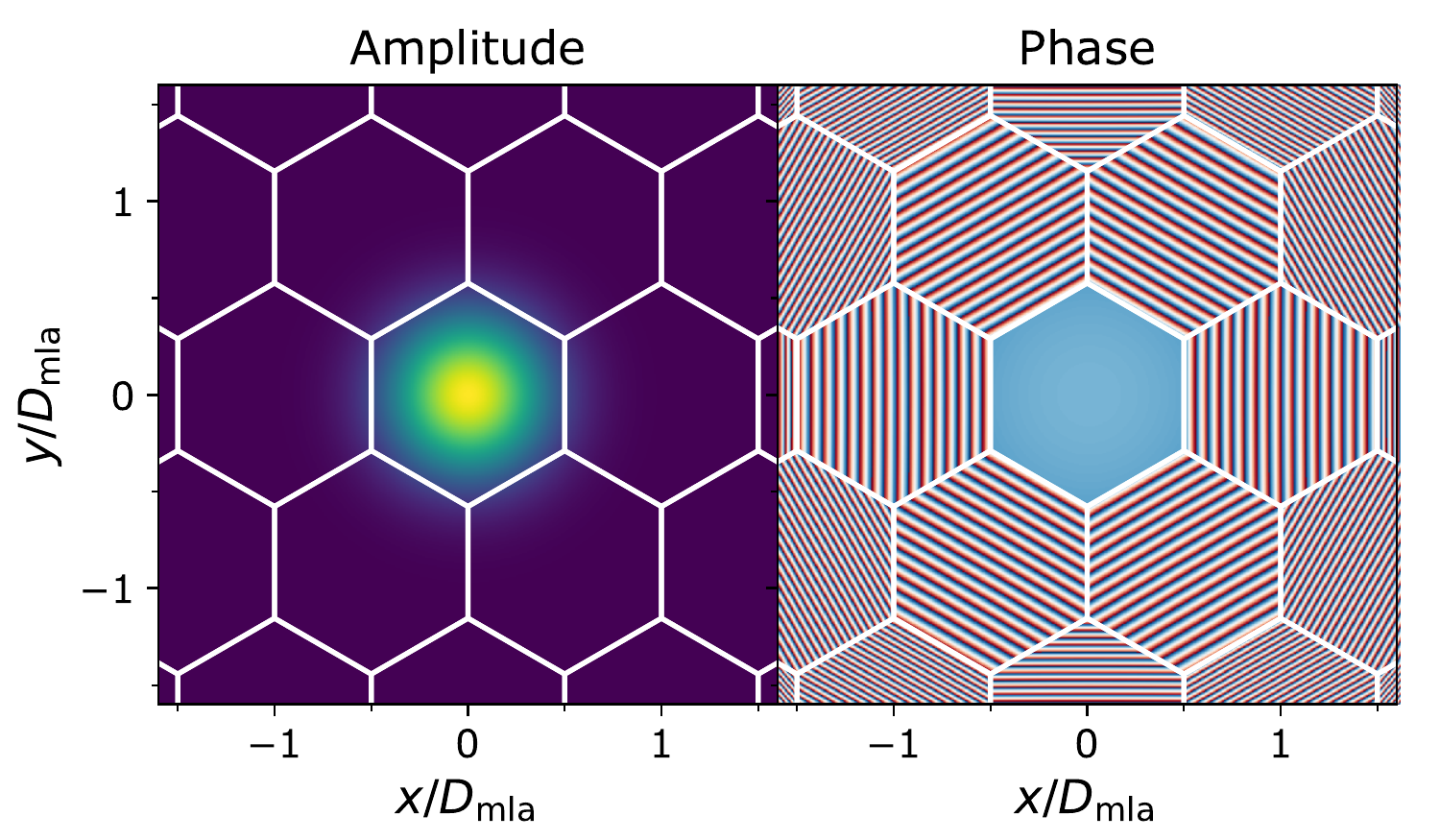}
\caption{Backpropagated mode of a single-mode fiber to the microlens array surface. Conceptually our microlens array and single-mode fiber can still be thought of as focal-plane electric field filtering using this modified mode. The mode is still Gaussian on the central microlens, but picks up an additional tilt on off-axis microlenses: on those microlenses the light needs to have a huge tilt to be propagated into the central fiber.}
\label{fig:mla_mode}
\end{figure}

\begin{figure}
\includegraphics[width=\columnwidth]{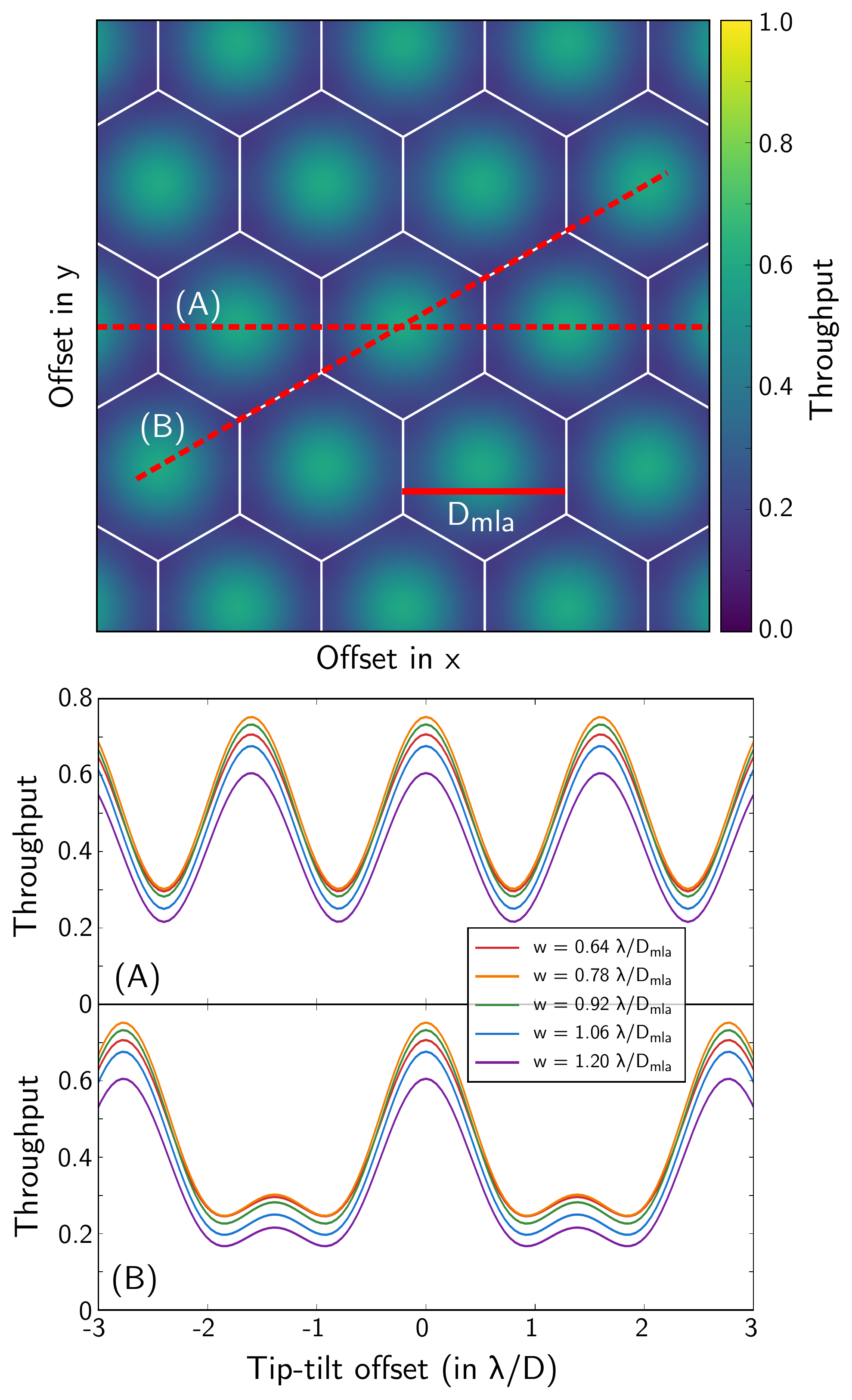}
\caption{The throughput of the single-mode fibers as a function of tip-tilt offset of the source. On the \changed{top} a two-dimensional throughput map is shown. On the \changed{bottom} two slices through this map are plotted for several values for the mode field diameter of the single-mode fibers. Maximum throughput of $\sim72\%$ is reached at the center of a lenslet. On the edge of two microlens the throughput of two fibers have to be added to reach $\sim30\%$ throughput. The worst case is the triple-point at which the maximum throughput drops to $\sim25\%$. A fiber mode-field radius of $w=0.78\lambda/D$ achieves the highest throughput for all tip-tilt offsets.}
\label{fig:throughput}
\end{figure}

To cover the field of view around a star, we need to fill the focal plane with single-mode fibers. This means that the fibers are impractically close together. A more reasonable solution is to use a microlens array with a single-mode fiber in each focus, as shown by \cite{corbett2009sampling}. Each fiber face now contains a strongly spatially-filtered telescope pupil. The corresponding focal-plane mode for each fiber can be recovered by back-propagating the fiber mode to the focal plane. An example of such a mode is shown in Fig.~\ref{fig:mla_mode}. The amplitude of this back-propagated mode is still Gaussian in amplitude. In phase however, it is flat within the central lenslet, but picks up a phase gradient on off-axis lenslets: light hitting off-axis lenslets need to have a huge tilt to still couple into the central fiber. We denote this off-axis contribution as lenslet crosstalk, and it is taken into account in every optimization and calculation done in this paper.

The throughput of the single-mode fiber array depends on the position of the object and the mode-field diameter of the fibers. Figure~\ref{fig:throughput} shows the throughput for a clear aperture with slices through the best and worst-case position angles. The throughput is dominated by the lenslet closest to the center of the PSF and is only weakly dependent on the mode-field diameter around the optimal value. Additionally, at each position in the focal plane, the optimal value of the mode-field diameter is the same, simplifying implementation.

\begin{figure}
\includegraphics[width=\columnwidth]{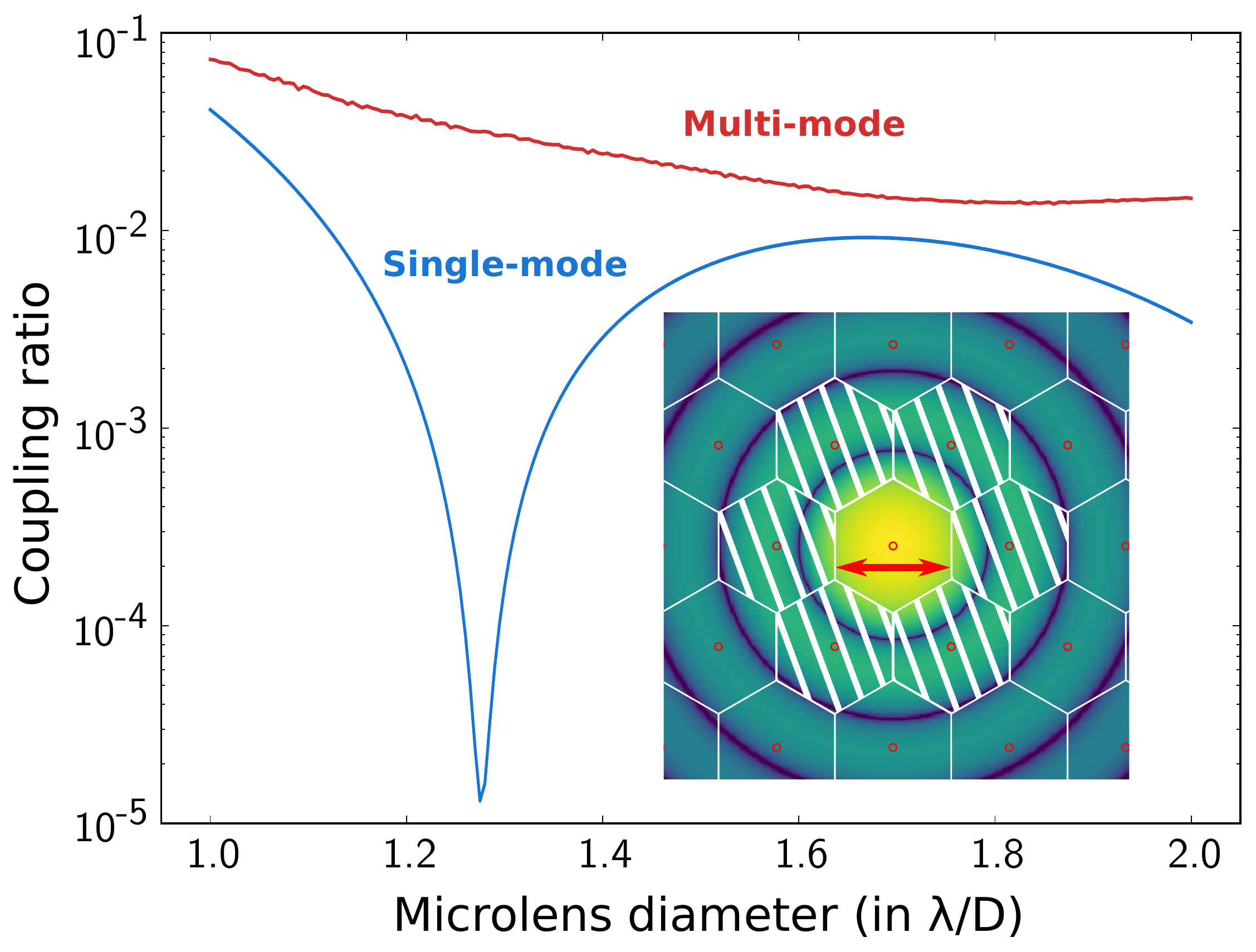}
\caption{\changed{Coupling ratio between an on-axis and off-axis source} through an off-axis microlens as a function of microlens diameter \changed{for a multi-mode and single-mode fiber}. The gain in contrast by using a single-mode fiber can readily be seen in the ratio of these two functions. For most microlens diameters this gain amounts to several orders of magnitude and reaches infinity at $\sim1.28\lambda/D$ where the light is perfectly nulled on the fiber face \changed{by the single-mode fiber}. This nulling is first order and is therefore very sensitive to wavelength and centering of the star around the central lenslet.}
\label{fig:mla_with_psf}
\end{figure}

Figure~\ref{fig:mla_with_psf} shows the throughput of an off-axis lenslet as a function of microlens diameter, while keeping the PSF centered around the on-axis lenslet. We can clearly see that at a diameter of $\sim1.28\lambda/D$ no stellar light is transmitted by the fiber. Note that this contrast is solely the result of the mode-filtering property of the single-mode fiber: if we were to use multi-mode fibers instead, the contrast would still be $\sim3\times10^{-2}$ at this point. The nulling can be classified as first order: only where the electric field of the Airy core and the first Airy ring exactly cancel do we see the contrast reduction. Moving the PSF only slightly already destroys this nulling.

Since the PSF changes in size with wavelength, the throughput of an off-axis fiber is inherently chromatic. We can read off the spectral bandwidth from Fig.~\ref{fig:mla_with_psf} directly. A contrast of $10^{-4}$ is reached for $1.26 \lambda/D < D_\mathrm{mla} < 1.30 \lambda/D$, corresponding to a spectral bandwidth of just $3\%$. Nevertheless this demonstrates that significant gains can be obtained by using single-mode fibers instead of multi-mode fibers or even conventional intensity detectors.

\section{Coronagraphy with a single-mode fiber array}
\label{sec:coronagraphy}

\subsection{Conventional coronagraphy}

We can use conventional coronagraphy techniques to reduce the spot intensity and ignore the mode-filtering property in the design process. As an example we use the apodizing phase plate (APP) coronagraph \citep{codona2006high,snik2012vector,otten2017sky}. This coronagraph consists of a single phase-only optic in the pupil plane, making it impervious to tip-tilt vibrations of the telescope or adaptive optics system. The phase pattern is designed to yield a dark zone in a certain region of interest in the focal plane. This region of interest can be both one- and two-sided, and can have arbitrary shapes. \changed{Most often the one-sided regions of interest are D-shaped and the two-sided are annular. See \cite{por2017optimal} for a recent description of APP design.} As both the PSF of the star and the planet are altered, the Strehl ratio is maximized to retain planet transmission.

\begin{figure}
\includegraphics[width=\columnwidth]{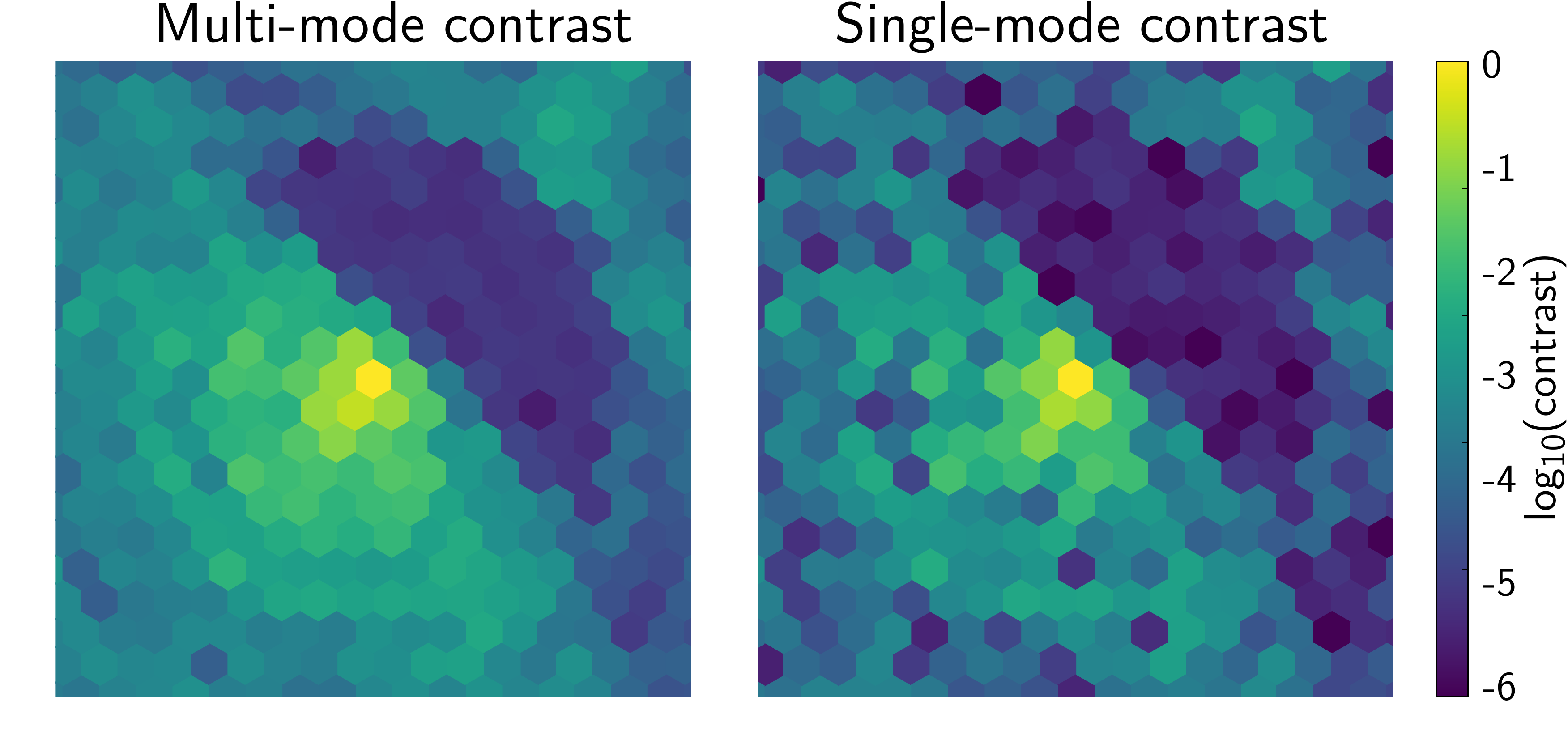}
\caption{Using a conventional coronagraph such as an APP contrast is still enhanced by the single-mode fibers. On the left is the raw contrast averaged over the microlens surfaces. On the right is the raw contrast through the single-mode fibers. The contrast in the dark zone is still enhanced by $\sim3\times$ on average.}
\label{fig:conventional_app}
\end{figure}

Figure~\ref{fig:conventional_app} shows the contrast through a fiber-fed single-mode fiber array using an APP designed for a contrast of $10^{-5}$ in a D-shaped region with an inner working angle of $2\lambda/D$ and outer working angle of $10\lambda/D$. While the use of single-mode fibers does enhance the contrast by $\sim3\times$ on average, this enhancement is not consistent: in some fibers the contrast is enhanced by $>10\times$ while in others we barely see any improvement at all. \changed{This shows that the factor of 3 enhancement that \cite{mawet2017observing} found for a dynamic random speckle field holds true for a single-mode fiber in a static structured speckle field, such as a residual coronagraphic electric field, even when the mode shape is modified from a Gaussian to a constricted Gaussian profile.}

\subsection{Direct pupil-plane phase mask optimization}
\comment{Aim: let coronagraph design be dependent on this mode filtering quality. This means that coronagraph design becomes easier, as they do not need the intensity to be zero, only the coupling through the fibers. Use APP coronagraph to engineer PSF to reach IWA.}

\begin{figure}
\centering
\includegraphics[width=\columnwidth]{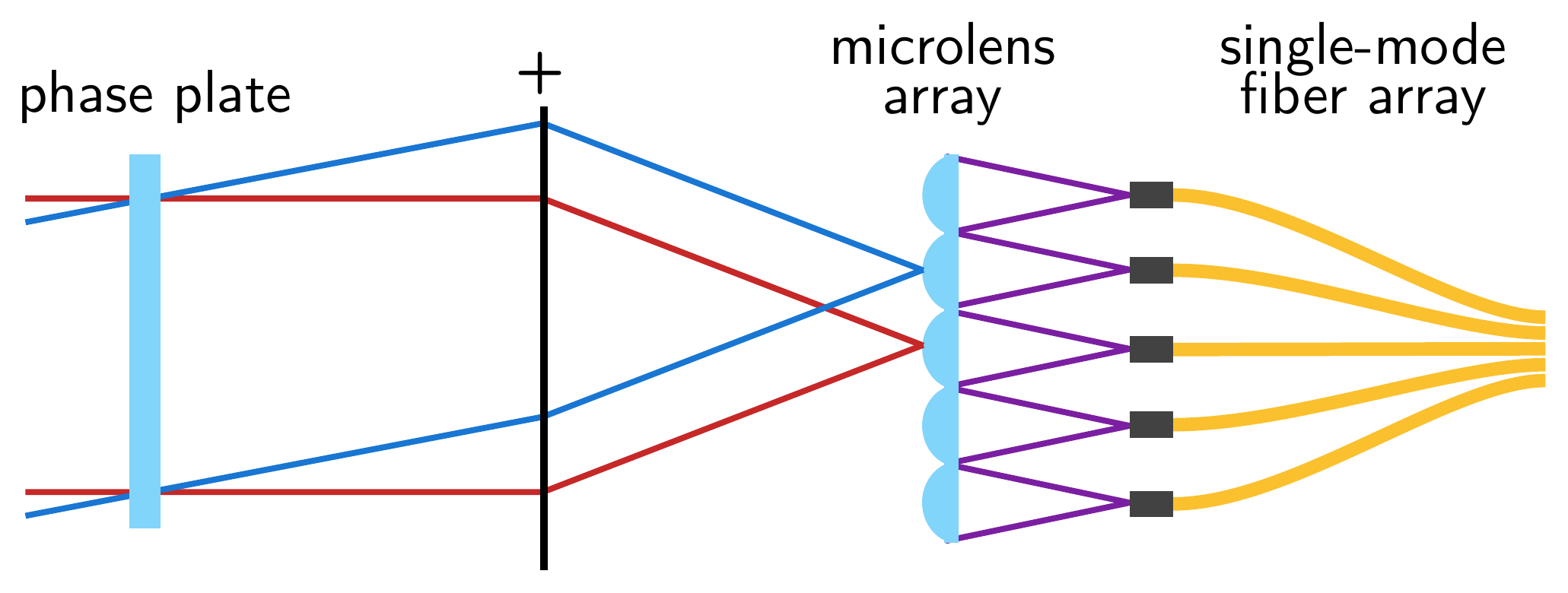}
\caption{The schematic layout of the proposed system. The phase plate located in the pupil-plane alters the PSF that is imaged on the microlens array. Each microlens focus is imaged on a single-mode fiber. An off-axis source will be spatially separated in the focal-plane and its Airy core will fall on a different microlens.}
\label{fig:schematic_layout}
\end{figure}

This improvement brings up the question: can we make use of this mode-filtering in the coronagraph design? As the single-mode fiber array already filters out some electric field modes, the coronagraph \changed{does not} have to suppress those modes; only modes that are transmitted by the single-mode fiber array need to be suppressed by the coronagraph. The coronagraph needs to minimize the coupling through the single-mode fibers, not the intensity at those positions in the focal plane. Designing a coronagraph specifically for the fiber array therefore \changed{allows for more design freedom compared to} conventional coronagraph design. In principle, any coronagraph can be designed to take the fiber coupling into account. As a case study, we use \changed{a pupil-plane phase plate} to alter the PSF in the focal plane. A schematic layout of the proposed system is shown in Fig.~\ref{fig:schematic_layout}.

\changed{To find the phase pattern,} we use the novel optimizer from \changed{\cite{por2017optimal}}, based on the work by~\cite{carlotti2013hybrid}, that maximizes the throughput (ie. Strehl ratio) for a complex pupil mask, while constraining the stellar intensity in the dark zone to be below the desired contrast. Since the transformation between the pupil and focal plane is linear in electric field, this optimization problem is linear, and its global optimum can be easily found using large-scale numerical optimizers such as Gurobi~\citep{gurobi}. In practice the optimization produces phase-only solutions, which is surprising as non-phase-only solutions are still feasible solutions. As the phase-only optimization problem is simply a more constrained version of the linear one, the phase-only solution must therefore be a global optimum of both problems.

\begin{figure*}[t!]
\centering
\includegraphics[width=\textwidth]{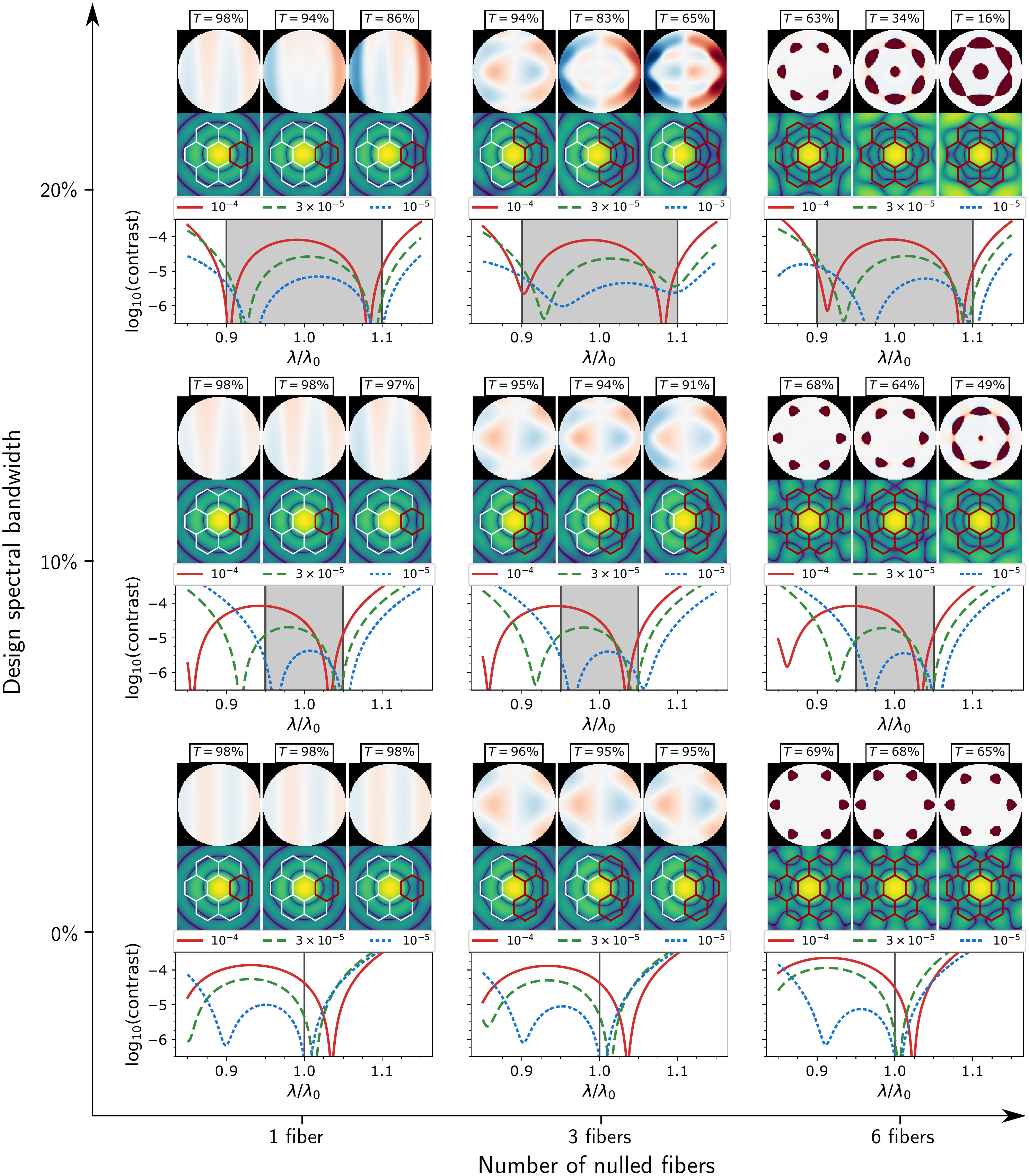}
\caption{A series of optimizations for 1, 3 and 6 fibers on the first ring of microlenses for a clear aperture. The design spectral bandwidth were 0\%, 10\% and 20\% and the contrasts $1\times10^{-4}$, $3\times10^{-5}$ and $1\times10^{-5}$. A $0.06\lambda/D$ peak-to-peak telescope tip-tilt jitter was also taken into account. \changed{Each microlens has a circum-diameter of $1.8\lambda/D$.} For each \changed{SCAR design} we show the \changed{pupil-plane} phase pattern, its \changed{corresponding} point spread function, its \changed{raw contrast $\eta_s/\eta_p$ as a function of wavelength averaged over the marked fibers} and its relative transmission compared to the unaltered PSF transmission. \changed{In this case, the unaltered PSF transmission was $78\%$ of the total light input. The relative transmission $T$ indicates the reduction in throughput due to the inclusion of the phase plate in the system.} The chromatic response shows the raw contrast after the single-mode fiber. The second-order nulling on the fiber face is clearly visible in every design. \changed{In Table~\ref{tab:design_parameters} we list the fixed and varied parameter in this figure. A summary of the throughput of all SCAR designs can be found in Figure~\ref{fig:throughput_matrix}.}}
\label{fig:app_clear}
\end{figure*}

\begin{figure*}[t!]
\centering
\includegraphics[width=\textwidth]{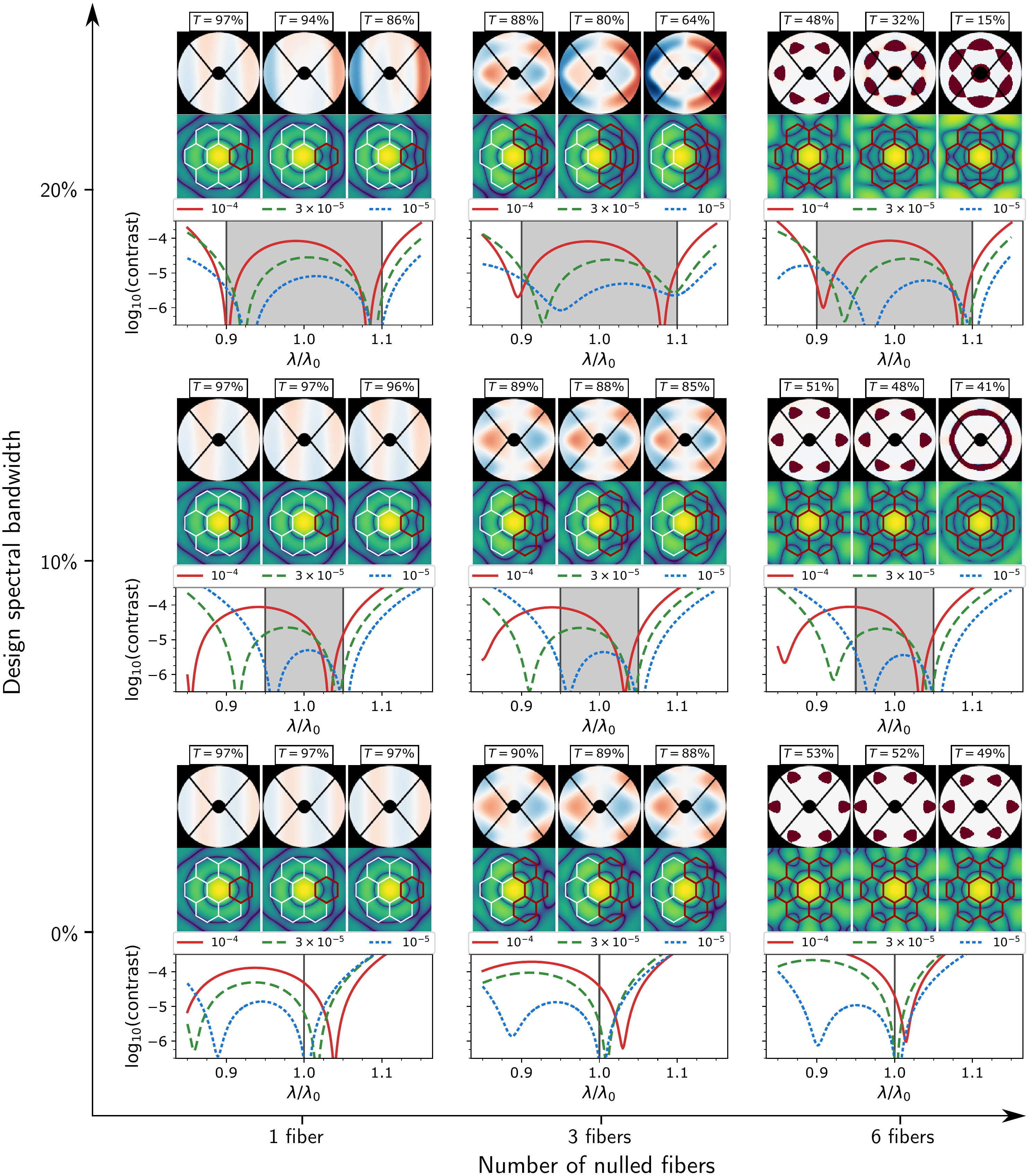}
\caption{The same as Fig.~\ref{fig:app_clear} for a VLT aperture. The aperture was subject to a 1\% binary erosion\changed{, ie. undersizing the pupil and oversizing central obscuration, spiders and other pupil features by 1\% of the aperture size,} to accommodate a pupil misalignment. The general structure of the solutions is similar to the case of a clear aperture. The central obscuration increases the strength of the first Airy ring, thereby decreasing the throughput of these \changed{SCAR designs} slightly. The relatively thin spiders have no influence on the throughput at these angular separations. \changed{The unaltered PSF transmission was $73\%$ of the total light input. In Table~\ref{tab:design_parameters} we list the fixed and varied parameter in this figure. A summary of the throughput of all SCAR designs shown in this figure can be found in Fig.~\ref{fig:throughput_matrix}.}}
\label{fig:app_vlt}
\end{figure*}

\begin{figure*}[t!]
\centering
\includegraphics[width=0.95\textwidth]{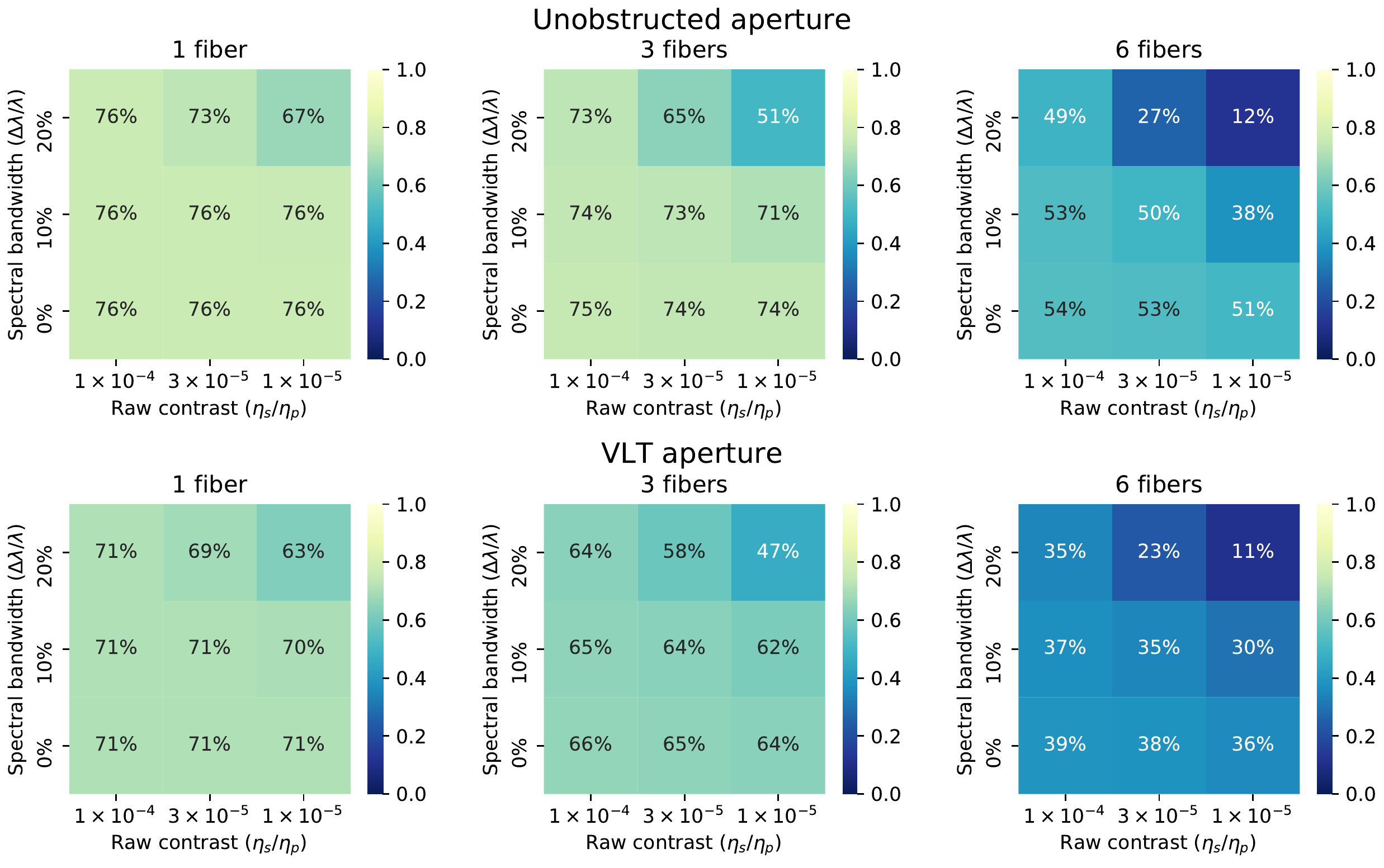}
\caption{\changed{The throughput ($\eta_p$) for the \changed{SCAR designs} shown Figs.~\ref{fig:app_clear}~and~\ref{fig:app_vlt}. The shown throughput is the total throughput of the coronagraphic system. It includes the Strehl reduction of the phase plate, the spatial filtering of the microlenses and the mode filtering of the single-mode fibers.}}
\label{fig:throughput_matrix}
\end{figure*}

The fiber coupling integral in Eq.\ref{eq:fiber_coupling}, or rather the amplitude of the coupled electric field 
\begin{equation}
E_\mathrm{coupled} = \frac{\int E_\mathrm{in}^* E_\mathrm{sm} \mathrm{d}A}{\left|\int E_\mathrm{sm} \mathrm{d}A\right|}
\end{equation}
is still linear in the input electric field $E_\mathrm{in}$, so we can apply the same method here. We maximize the throughput of the central fiber, while constraining the coupling through the specified off-axis fibers. To counter the chromaticity mentioned in Sect.~\ref{sec:single_mode_fiber_array}, we constrain the off-axis stellar intensity at several wavelengths simultaneously, which ensures that the contrast is attained over a broad spectral bandwidth. Jitter resistance is kept in check in a similar manner: the desired \changed{raw} contrast must be attained for several tip-tilt positions simultaneously.

In Fig.~\ref{fig:app_clear} we show a few examples of optimizations for 1, 3 and 6 fibers for a contrast of $1\times10^{-4}$, $3\times10^{-5}$ and $1\times10^{-5}$ using a $0\%, 10\%$ and $20\%$ spectral bandwidth, along with their corresponding PSF and chromatic response. \changed{The design parameters are shown in Table~\ref{tab:design_parameters}. These design parameters were chosen to show a variety of SCAR designs for realistic implementations. At the shown contrast limits, the residual atmospheric speckles will limited the on-sky contrast, even after an extreme AO system. The spectral bandwidths were chosen as wide as possible, without compromising on planet throughput. The resulting spectral bandwidths are large enough to apply spectral cross-correlation techniques.}

\begin{table}
\centering
\caption{\changed{The design parameters used for all SCAR designs throughout this work. All SCAR designs generated with these parameters can be found in Figures~\ref{fig:app_clear}~and~\ref{fig:app_vlt}.}}
\label{tab:design_parameters}
\begin{tabular}{ll}
\hline\hline
Parameter name & Value \\ \hline
Raw contrast limit & $\{1\times10^{-4}, 3\times10^{-5}, 1\times10^{-5}\}$ \\
Spectral bandwidths & $\{0\%, 10\%, 20\%\}$ \\
Tip-tilt errors & $0.06\lambda/D$ peak-to-peak \\
Microlens circum-diameter & $1.8 \lambda/D$ \\
Microlens shape & Hexagonal \\
Fiber mode shape & Gaussian \\
Fiber mode-field-diameter & $1.7 \lambda/D_\mathrm{mla}$ \\
Pupil mask & $\{\mathrm{unobstructed}, \mathrm{VLT}\}$ \\ \hline
\end{tabular}
\end{table}

Note that in each case the optimizer prefers a second order null. This second order null is much more stable against bandwidth and tip-tilt jitter. The reason for this is explained graphically in Fig.~\ref{fig:principle}. Furthermore, note that this second order null is even present in monochromatic optimizations and in optimizations without tip-tilt errors added. This means that the second-order null requires less phase stroke to achieve and therefore provides higher Strehl ratios.

As the optimizer can handle arbitrary apertures, optimizations for other aperture shapes are also possible. Figure~\ref{fig:app_vlt} shows optimizations for a VLT aperture for the same parameters as for the clear apertures. The aperture was subject to a 1\% binary erosion\changed{, ie. undersizing the pupil and oversizing central obscuration, spiders and other pupil features by 1\% of the aperture size,} to accommodate for a misalignment in the pupil mask. Although the overall structure is quite similar, there is one key difference \changed{compared to a clear aperture: the relative transmission $T$ is lower for all phase plate designs.} \changed{This means that the relative transmission} depends strongly on the size of the central obscuration. This is obvious as larger central obscurations strengthen the first Airy ring and brighter features typically cost more stroke, and therefore \changed{relative transmission}, to change, \changed{similar to conventional APP design}. Effectively, this means that each feature in the \changed{phase} pattern becomes larger to compensate for larger central obscurations.

\changed{We summarize the multitude of SCAR phase pattern designs in Figure~\ref{fig:throughput_matrix}. This figure shows the total planet throughput $\eta_p$, provided that the planet is located in the center of the off-axis microlens. This throughput includes all theoretically unavoidable terms, but excludes all experimental terms. A summary of important throughput terms are listed in Table~\ref{tab:throughput_terms}.}

\begin{table}
\centering
\caption{\changed{The different throughput terms that are important for the SCAR coronagraph. A distinction is made between theoretical and experimental terms. Experimental throughput terms will be non-existent with perfect manufacturing, while theoretical throughput terms are unavoidable. Typical values in the visible are shown for each term.}}
\label{tab:throughput_terms}
\begin{tabular}{ll}
\hline\hline
Throughput term & Typical values \\
\hline
\textbf{Theoretical} \\
~~~~~Geometric lenslet throughput & $\sim80\%$ \\
~~~~~Fiber injection losses & $90\%-95\%$ \\
~~~~~Planet location & $50\%-100\%$ \\
~~~~~Phase plate Strehl reduction & $60\%-80\%$ \\
\textbf{Experimental} \\
~~~~~Phase plate transmission & $>85\%$ \\
~~~~~Fresnel losses on the fiber & $\sim90\%$ \\
~~~~~Microlens transmission & $>95\%$ \\
~~~~~Strehl ratio of the AO system & $\sim50\%$ \\
\hline
\end{tabular}
\end{table}

In the rest of this paper, we consider the outlined \changed{SCAR design} in Fig.~\ref{fig:app_clear}~and~\ref{fig:app_vlt} using a 10\% spectral bandwidth for a contrast of $3\times10^{-5}$. Although optimized for only 10\%, \changed{this specific design performs exceptionally well and} a contrast of $<10^{-4}$ is obtained for a spectral bandwidth of 18\% centered around the design wavelength.

\section{Single-mode fiber coronagraph properties}
\label{sec:properties}

\comment{Introduce clear APP and VLT APP design that will be used for all properties}

In this section we \changed{show} the properties of this new coronagraph \changed{and perform parameter studies on the fixed parameters in Table~\ref{tab:design_parameters}. We discuss the mode-field-diameter of the single-mode fiber in Sect.~\ref{sec:fiber_mfd},} throughput and inner-working angle in Sect.~\ref{sec:throughput}, the chromatic response in Sect.~\ref{sec:chromaticity}, the tip-tilt sensitivity of the \changed{SCAR} designs in Sect.~\ref{sec:tip-tilt} and the sensitivity of other modes in Sect.~\ref{sec:modal_sensitivity}.

\subsection{Fiber mode-field-diameter}
\label{sec:fiber_mfd}

\begin{figure}
\includegraphics[width=\columnwidth]{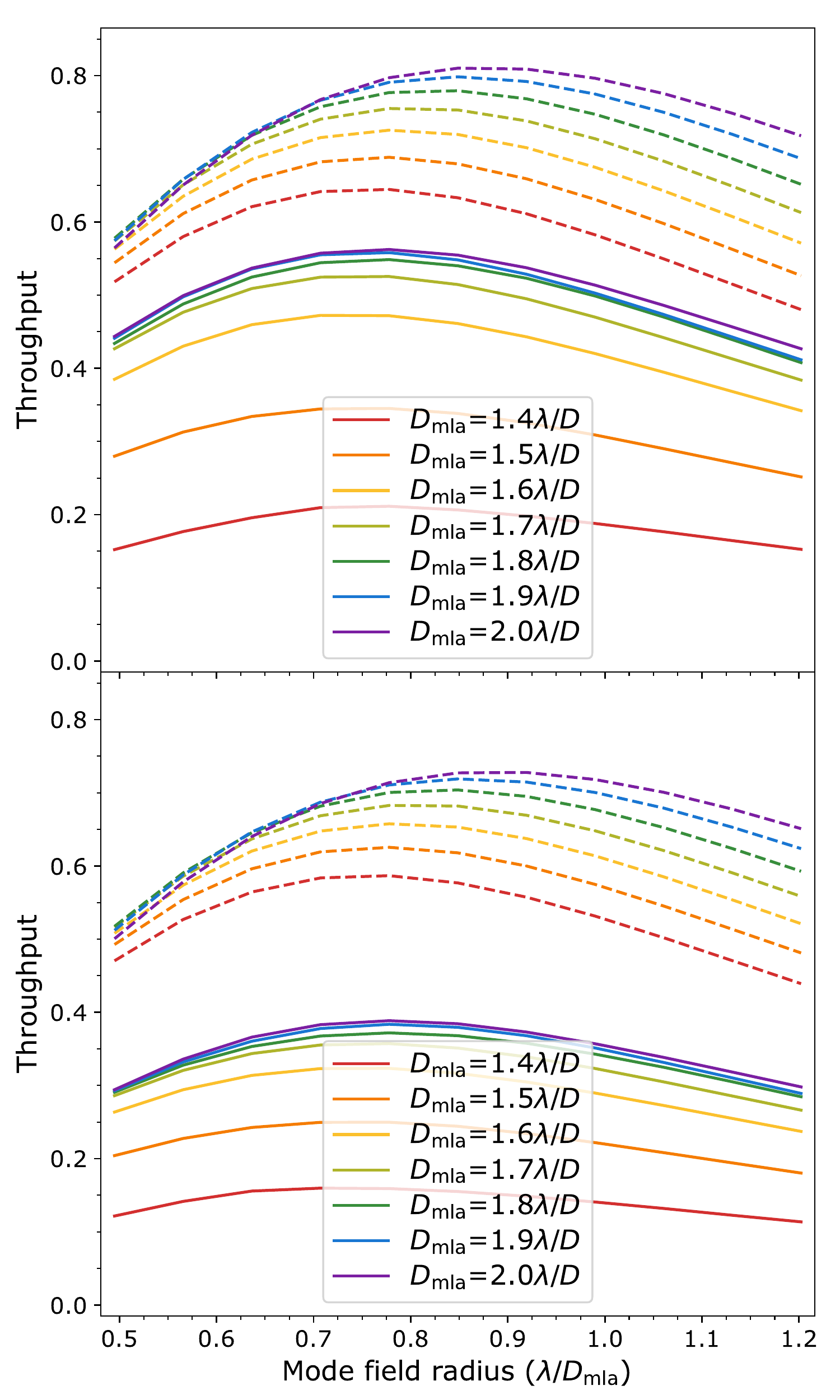}
\caption{The throughput of the central lenslet as a function of mode field radius for various values of the microlens diameter. Each data point represents a different \changed{SCAR design}. Solid lines indicate the \changed{SCAR PSFs}, dashed lines an unaltered PSF. The top panel shows the throughput for an clear aperture, the bottom panel for a VLT aperture.}
\label{fig:mfd_throughput}
\end{figure}

The \changed{phase plate} reduces the throughput of planet light. This reduction \changed{will also} affect the optimal value of the mode-field diameter. Smaller mode-field diameters result in larger back-propagated fiber modes in the focal plane, which makes it easier to squeeze the three rings necessary for the second-order nulling into this mode. Therefore, we expect higher Strehl ratios (the throughput relative to the unaltered PSF throughput) as the mode-field diameter becomes smaller. This is superimposed on the actual throughput of the unaltered PSF. Both curves are shown in Fig.~\ref{fig:mfd_throughput} for both a clear and the VLT aperture.

\subsection{Throughput and inner-working angle}
\label{sec:throughput}

The throughput shown here is the fractional transmission of light from the entire pupil into the central single-mode fiber: it includes \changed{all theoretical terms as listed in Table~\ref{tab:design_parameters}. It however excludes all experimental terms.} It is clear that larger microlenses generally give a better throughput, as is expected. \changed{Additionally, we can see that the} optimal mode-field diameter as a function of microlens diameter for the unaltered PSF moves to larger mode-field diameters, as it is essentially matching the Airy-core width rather than the size of the microlens itself. The optimal mode-field diameter for the \changed{SCAR} however stay the same as smaller mode-field diameter have an advantage in their Strehl ratio.

\begin{figure}
\centering
\includegraphics[width=\columnwidth]{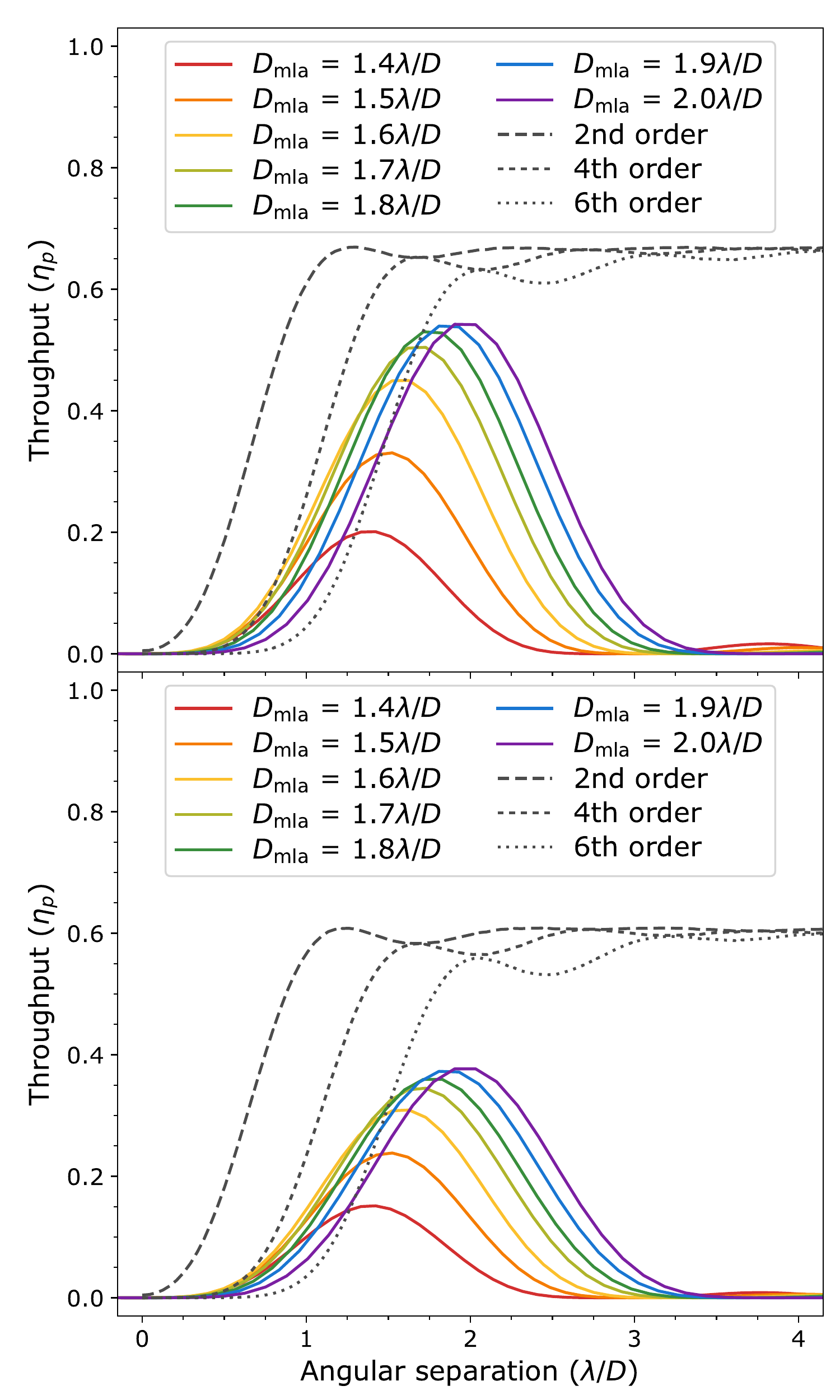}
\caption{The throughput as function of off-axis angle for various values of the microlens diameter. The throughput of a perfect second, fourth and sixth-order coronagraph is also plotted. The top panel shows the throughput for an unobstructed aperture, the bottom panel for the VLT aperture. \changed{The throughput of the theoretical coronagraphs is taken to be the fractional flux within an aperture of radius $0.7\lambda/D$ centered around the planet.}}
\label{fig:useful_throughput}
\end{figure}

Figure~\ref{fig:useful_throughput} shows the throughput \changed{($\eta_p$)} of the coronagraph for different values for the microlens diameter. The mode-field radius \changed{of the fiber} was fixed at $w=0.85\lambda/D_\mathrm{mla}$ and the contrast at $3\times10^{-5}$. We adopted a $0.1\lambda/D$ rms telescope tip-tilt jitter with a normal distribution, \changed{corresponding to a 2 mas rms tip-tilt jitter at a wavelength of $\lambda=750$nm. This level of tip-tilt jitter was chosen to mimic a SPHERE-like adaptive optics system, according to \cite{fusco2016saxo}}. The throughput is averaged over all pointing positions and over the full 10\% \changed{spectral} bandwidth. The throughput of off-center fibers is negligible to that of the central fiber: all throughput is concentrated in only one single-mode fiber.

As the stellar throughput is only minimized for the first ring of microlenses, we do not have much throughput beyond the edge of the first off-axis microlens. \changed{Designs can be made for more than one ring of microlenses, although this complicates the design procedure and will be discussed in future work.} Extremely close to the star, we have no throughput, as the Airy core is still mostly on the central lenslet. At $\sim0.5\lambda/D$ the throughput starts to rise, reaching a maximum at the center of the first microlens. Note that a throughput of $\sim50\%$ of the maximum \changed{SCAR} throughput is already reached at $\sim1\lambda/D$, \changed{which is the usual definition of inner working angle}. Also note that up to microlens diameters of $\sim1.8\lambda/D$ the throughput at small angular separations \changed{($<1\lambda/D$)} does not change much, but the maximum throughput still increases. For larger microlens diameter we still gain in throughput at the center of the microlens, however the throughput at these small angular separations starts to suffer, \changed{which is especially visible in the $D_\mathrm{mla}=2.0\lambda/D$ throughput curve.}

Lines for the theoretical throughput of other coronagraphs are overplotted in \changed{Fig.}~\ref{fig:useful_throughput}. Perfect coronagraphs refer to the notion introduced by~\cite{cavarroc2006fundamental} and \cite{guyon2006theoretical}. A second-order perfect coronagraph removes a constant term from the pupil-plane electric field. A fourth-order additionally removes the $x$ and $y$ components from the electric field. A sixth-order perfect coronagraph furthermore removes the $x^2$, $xy$ and $y^2$ modes from the electric field. \changed{For the theoretical coronagraphs, the throughput is calculated using a circular aperture of $0.7\lambda/D$ centered around the off-axis planet. We can see that the SCAR throughput lags behind the theoretical second-order coronagraph, but stays close to the fourth-order and beats the sixth-order at angular separations $<1.7\lambda/D$.}

\begin{figure}
\centering
\includegraphics[width=\columnwidth]{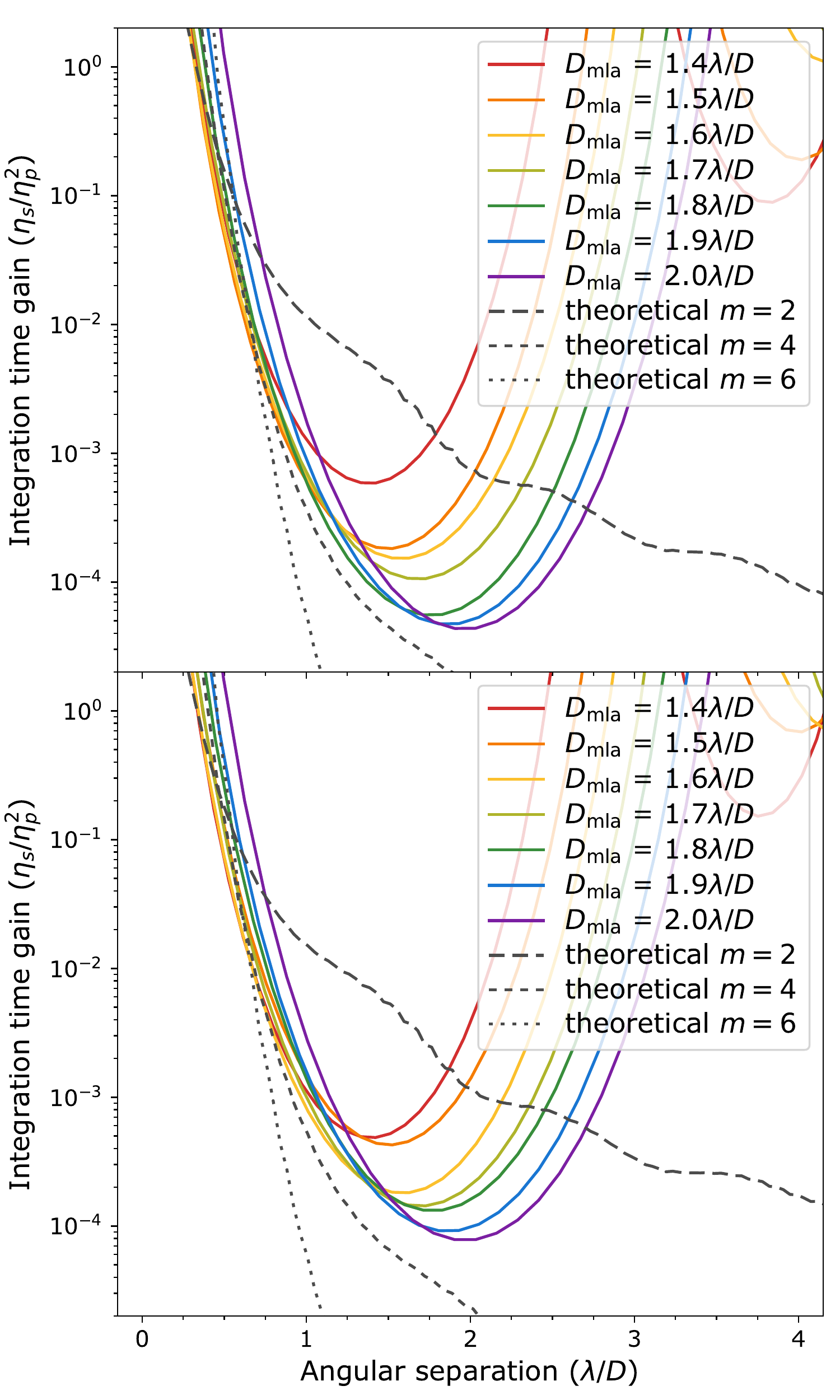}
\caption{\changed{The integration time gain ($\eta_s/\eta_p^2$) as function of off-axis angle for various values of the microlens diameter. The top panel shows the integration time gain for an unobstructed aperture, the bottom panel for the VLT aperture. The integration time for the theoretical coronagraphs is calculated on the flux within an aperture of radius $0.7\lambda/D$ centered around the planet.}}
\label{fig:integration_time_gain}
\end{figure}

\changed{Figure~\ref{fig:integration_time_gain} shows the integration time gain ($\eta_s/\eta_p^2$) under the same conditions as in Fig.~\ref{fig:useful_throughput}. We now see that, even though the throughput of the theoretical second-order coronagraph is good, its integration time is minor because it doesn't outweigh the loss in starlight suppression. SCAR however performs similar to theoretical fourth-order coronagraph for angular separations $<1.8\lambda/D$. A sixth-order coronagraphs does even better, but suffers from a lack of throughput which becomes noticeable in cases where the raw contrast ($\eta_s/\eta_p$) is limited, which is the case in any ground-based telescope. This suggests that SCAR is a close-to-optimal coronagraph.}

\subsection{Spectral bandwidth}
\label{sec:chromaticity}

Figures~\ref{fig:app_clear}~and~\ref{fig:app_vlt} show the chromatic response for all designs. Every design exhibits the double-dipped structure of the second-order null on the fiber. For all designs with a non-zero spectral bandwidth, we can also see that the contrast is hard to achieve on the long wavelength side. At these longer wavelengths the bright Airy core starts to grow into the microlens array. This means that the second Airy ring needs to be made much brighter to compensate, which requires substantial deviations in the phase pattern. \changed{Qualitatively,} the location of the second null is chosen \changed{by the optimizer} such that the spectral bandwidth requirement is reached.

\subsection{Tip-tilt sensitivity}
\label{sec:tip-tilt}

\begin{figure}
\includegraphics[width=\columnwidth]{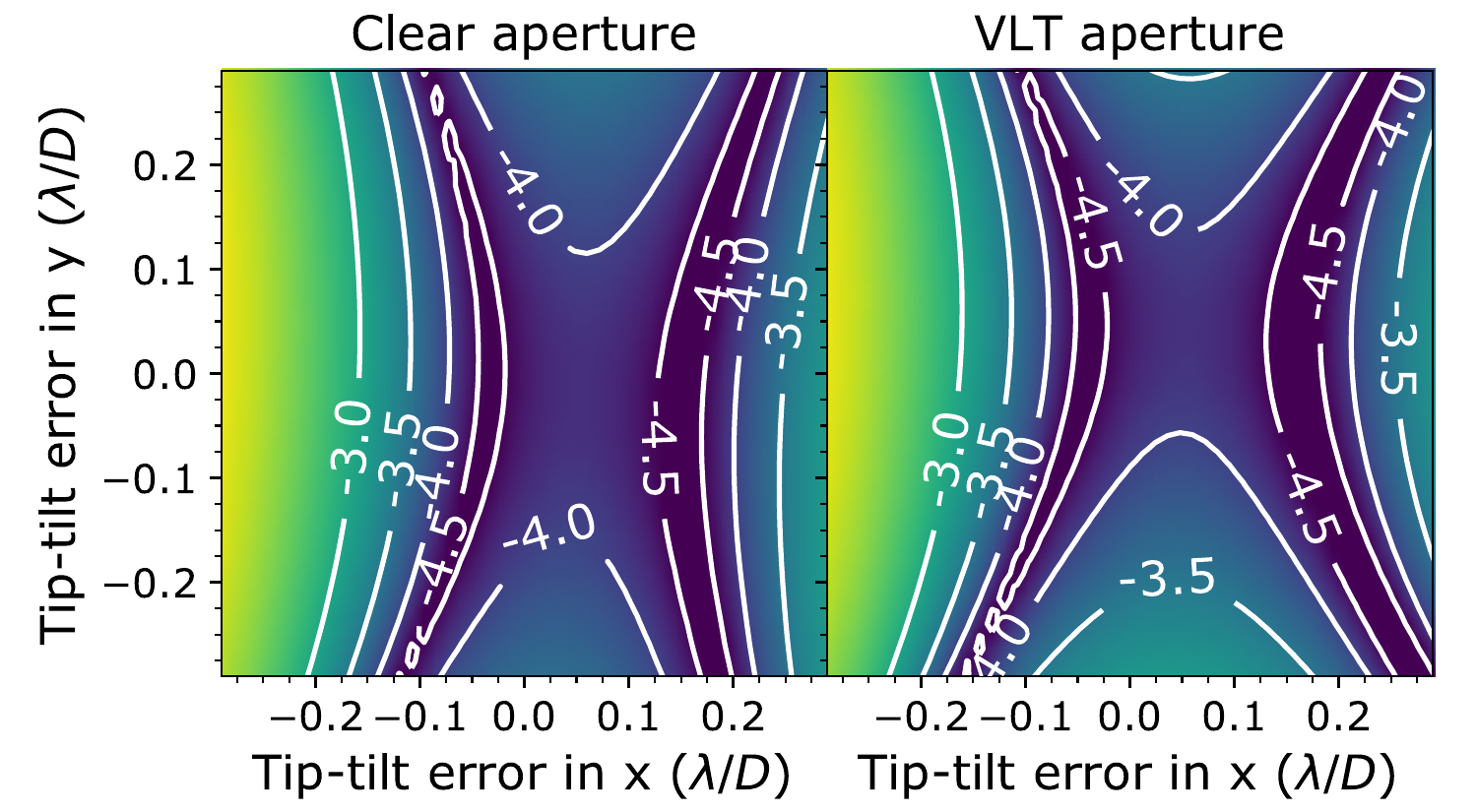}
\caption{Map of the worst contrast through an off-axis fiber over a 10\% bandwidth as a function of tip-tilt error upstream of the microlens array. This shows that the coronagraph is reasonably stable against tip-tilt, allowing for a $\sim0.15 \lambda/D$ tip-tilt error until the contrast drops to $10^{-4}$. \changed{The contour labels indicate $\log_{10}(\eta_s/\eta_p)$.}}
\label{fig:vibration_sensitivity}
\end{figure}

Figure~\ref{fig:vibration_sensitivity} shows the average contrast \changed{($\eta_s/\eta_p$)} over the full 10\% bandwidth as a function of tip-tilt error upstream of the fiber injection unit. The double-dip structure is again clearly visible, which greatly improves the tip-tilt response. Both coronagraph designs achieve a tip-tilt stability of $\sim0.1\lambda/D$ rms.

\subsection{Sensitivity to other aberrations}
\label{sec:modal_sensitivity}

To show the sensitivity to other aberrations, we perform a sensitivity analysis on the SCAR coronagraph: we aim to find the mode basis of orthogonal modes ordered by their sensitivity. These principal modes can be found by taking the first-order Taylor expansion of the phase in the pupil-plane around the nominal position. In this way a linear transformation $G_\lambda$ can be constructed from a phase deformation $\delta \varphi$ to the resulting change in electric field in the fibers $\delta E_\lambda$:
\begin{equation}
\delta E_\lambda = G_\lambda \delta \varphi.
\end{equation}
Note that $G_\lambda$ and $\delta E_\lambda$ both depend on wavelength as the response of the coronagraph is inherently chromatic. A singular value decomposition of the matrix $G_\lambda$ yields the monochromatic principal phase modes of the coronagraph. The corresponding singular values denote the importance of those modes. Note that this expansion is similar to the one used in electric field conjugation~\citep{give2007broadband}.

Broadband principal modes can be obtained by stacking several $G_\lambda$ matrices for wavelengths within the spectral bandwidth into a single matrix $G$ as
\begin{equation}
\begin{bmatrix}
\delta E_{\lambda_1} \\
\delta E_{\lambda_2} \\
\vdots \\
\delta E_{\lambda_N} \\
\end{bmatrix}
=
\begin{bmatrix}
G_{\lambda_1} \\
G_{\lambda_2} \\
\vdots \\
G_{\lambda_N} \\
\end{bmatrix}
\delta\varphi.
\end{equation}
A singular value decomposition on the matrix $G$ now yields the broadband principal modes. The singular values are now indicative of the amount of electric field each phase mode induces in the fibers as a function of wavelength. Note that this method is again similar to the one used in broadband electric field conjugation~\citep{give2007broadband}.

\begin{figure}
\includegraphics[width=\columnwidth]{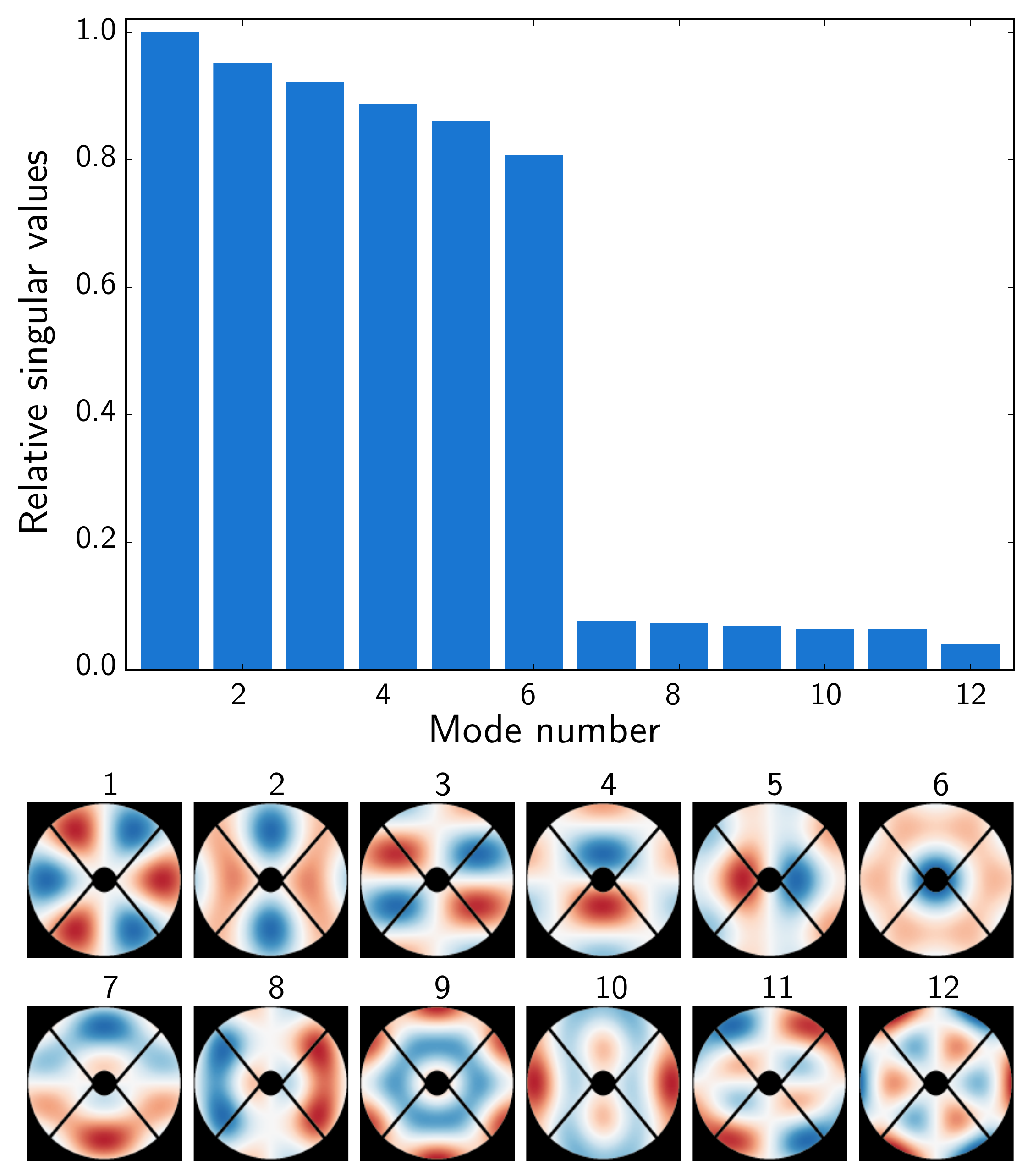}
\caption{Principal phase modes for the SCAR coronagraph using the \changed{design} described in the text. The top panel shows the singular value of each mode, indicating its significance for the obtained contrast after phase correction. The bottom panel shows the pupil-plane phase for each mode.}
\label{fig:principal_modes}
\end{figure}

Figure~\ref{fig:principal_modes} shows the broadband principal modes for the \changed{SCAR} design for the VLT aperture, along with their singular values. Only 6 modes are important for the final contrast. Naively we would expect 2 modes per fiber, so 12 modes in total, as we need to control both the real and imaginary part of the electric field. However in our case the system shows an anti-Hermitian symmetry: the transmitted electric field on fibers in opposite points in the focal plane are not independent if only small phase aberrations are present. This means that one phase mode determines the electric field for both fibers so that only 2 modes per 2 fibers are needed. Only half of the original 12 modes determine the contrast in a monochromatic system, meaning that 6 modes are left. The omitted 6 modes correspond to amplitude errors. For the broadband principal modes we of course expect some additional modes with low importance, corresponding to the spectral bandwidth increase. The first 6 of these are shown as modes 7 to 12 in Fig.~\ref{fig:principal_modes}.

The first principal mode is trefoil, which is the result of the six-fold symmetry in the fiber locations. Note that the other trefoil is missing: its corresponding electric field change is completely filtered out by the single-mode fibers: it creates radial first-order nulls directly on the center of every lenslet. Other important modes include secondary-astigmatism-like modes for modes 2 and 3, coma-like modes for modes 4 and 5, and a perturbed spherical mode for mode 6.

\section{Comparison to the vortex coronagraph}

\begin{figure}
\includegraphics[width=\columnwidth]{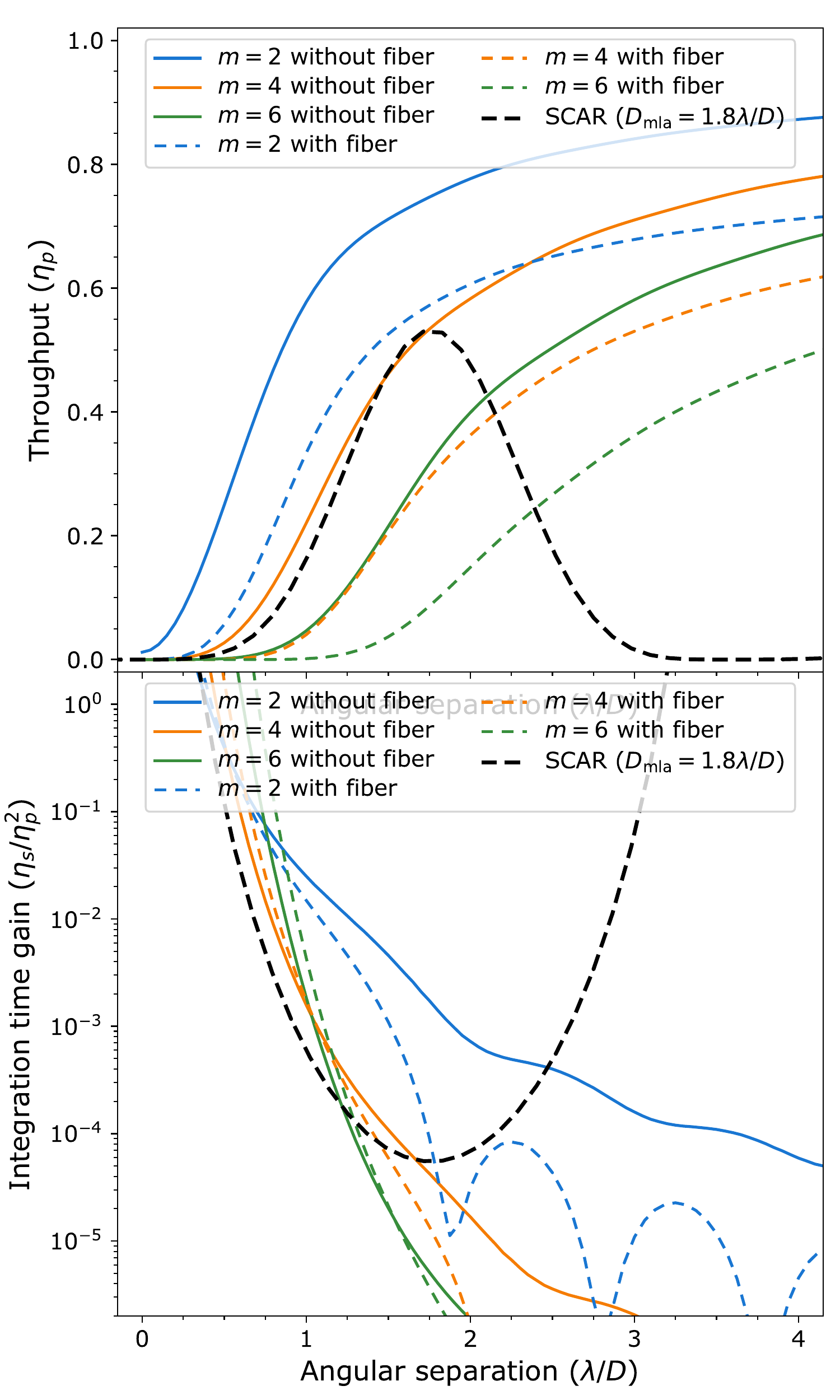}
\caption{The \changed{throughput ($\eta_p$) and integration time gain ($\eta_s/\eta_p^2$)} for a vortex coronagraph using a clear aperture for charges $m=2,4,6$ using a conventional intensity detector or a single-mode fiber centered around the planet. A telescope tip-tilt jitter of $0.1\lambda/D$ rms was taken into account. \changed{For the vortex coronagraph without fiber, the throughput and integration time gain is calculated on an aperture of radius $0.7\lambda/D$ centered around the planet.}}
\label{fig:useful_throughput_vortex}
\end{figure}

The described performance of \changed{the SCAR} coronagraph begs the question on how \changed{it compares to} other coronagraphs using single-mode fibers. \changed{In this section we will provide a comparison of the SCAR coronagraph with the single-mode fiber injection unit with vortex coronagraph proposed by \cite{mawet2017observing}, and a comparison with a multiplexed fiber injection unit as shown in Sect.~\ref{sec:coronagraphy} behind a vortex coronagraph.} Figure~\ref{fig:useful_throughput_vortex} shows the throughput \changed{and integration time gain} of a conventional vortex coronagraph using a clear aperture. A conventional intensity detector \changed{(ie. sum of all flux in an aperture centered around the star)} is compared with a single-mode fiber centered around the planet. The mode-field diameter was optimized for maximum throughput of the PSF without a coronagraph. Similarly to Fig.~\ref{fig:useful_throughput}, a telescope tip-tilt jitter of $0.1\lambda/D$ rms was chosen.

For charges $m=4$ and $m=6$ a decrease in throughput, \changed{compared to the vortex coronagraph with multi-mode fiber,} can be seen, approximately corresponding with the maximum coupling of an Airy pattern through a Gaussian single-mode fiber. Additionally the vortex in the focal plane imprints a phase ramp on off-axis sources, which degrades throughput even further. \changed{This effect is more pronounced with smaller angular separations and higher charge vortices.} Note that \changed{most of} this phase ramp can be easily negated by tilting the fiber slightly, depending on the focal-plane position of the planet. \changed{We can see that the SCAR coronagraph wins in throughput compared to vortex coronagraphs with charger $m>2$. However the charge 2 vortex coronagraph doesn't suppress the star very well, resulting in a moderate integration time gain compared to the SCAR coronagraph.}

\begin{figure}
\includegraphics[width=\columnwidth]{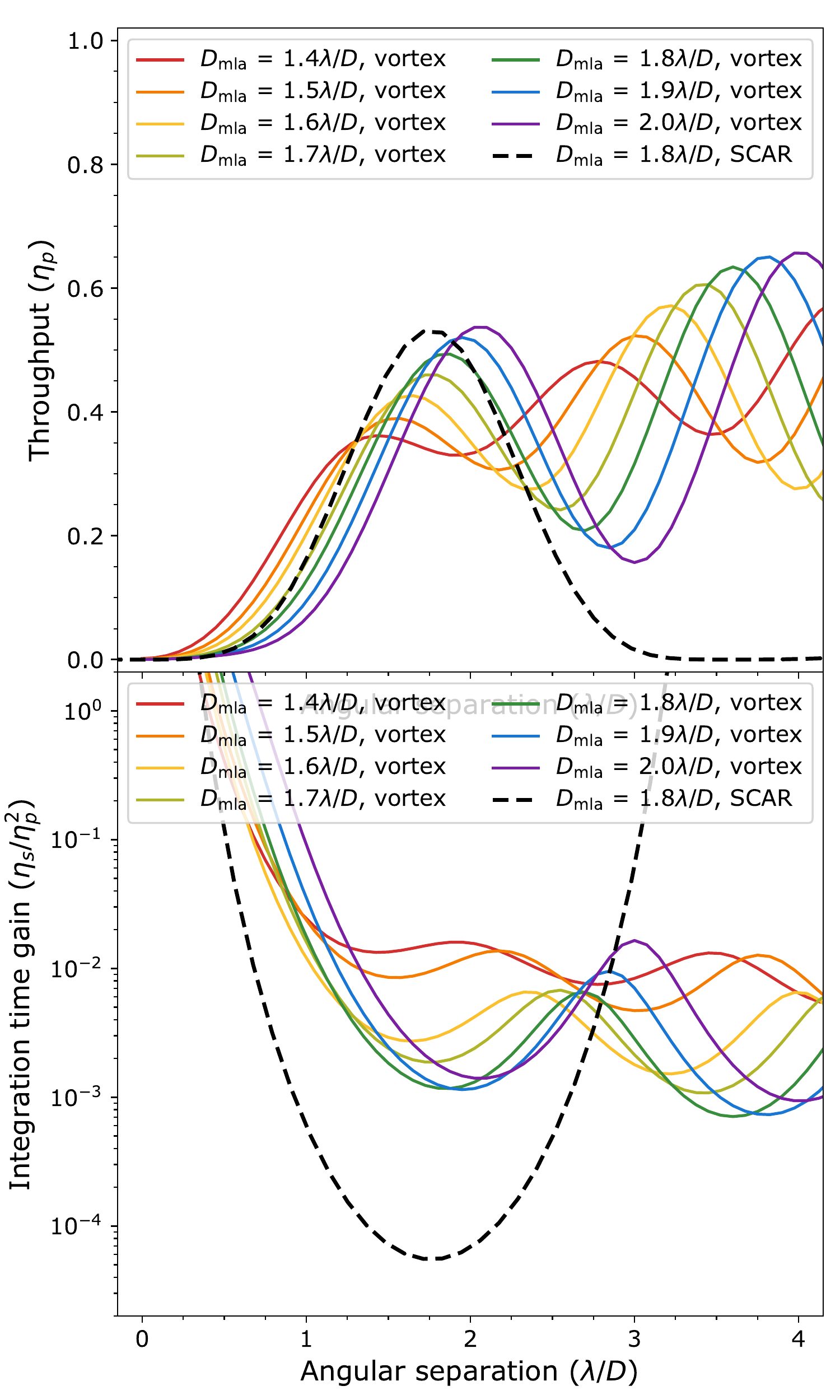}
\caption{\changed{The throughput ($\eta_p$) and integration time gain ($\eta_s/\eta_p^2$) of a charge $m=2$ vortex coronagraph on a microlens-fed single-mode fiber array. The diameter of the microlenses is varied from $1.4$ to $2.0\lambda/D$, and the mode-field diameter is optimized for maximum throughput of an Airy pattern. The throughput of the SCAR coronagraph designed for $1.8\lambda/D$ microlens diameter was added for comparison.}}
\label{fig:useful_throughput_vortex_mla}
\end{figure}

Figure~\ref{fig:useful_throughput_vortex_mla} shows the throughput \changed{and integration time gain} of a vortex coronagraph charge $m=2$ through a microlens-fed single-mode fiber-array as the fiber injection unit. The diameter of the microlenses is varied from $1.4$ to $2.0\lambda/D$, and the mode-field diameter is optimized for maximum throughput of an Airy pattern. For large angular separations the throughput oscillates due to the transmission of the microlens array as shown in Fig.~\ref{fig:throughput}. For the first ring of microlenses the throughput rises quickly, again reaching its maximum at the center of the microlens. Note that even though the \changed{coupling efficiency} for smaller microlenses is \changed{higher}, the \changed{geometric} throughput \changed{decreases more rapidly. A trade-off between these two throughput terms} leads to an optimal microlens diameter of $\sim1.8\lambda/D$ for the vortex coronagraph as well. The throughput at this microlens diameter is comparable to the performance of the SCAR coronagraph even though the vortex coronagraph has a more complicated optical setup. \changed{This multiplexed single-mode fiber vortex coronagraph however does not suppress starlight as well as the SCAR, leading to a worse integration time gain.} The vortex coronagraph however does have the advantage of an infinite outer-working angle.

\section{Conclusion}
\label{sec:conclusion}

In this paper we described the principle behind \changed{coronagraphs leveraging the design freedom offered by the use of single-mode fibers as a mode filter. We have shown the properties of a microlens-array fed single-mode fiber-array, making it possible to perform exoplanet searches.} We combined this with a \changed{pupil-plane phase plate, yielding the SCAR coronagraph} and presented the following advantages \changed{of this new coronagraph}:
\begin{enumerate}
\item \emph{Low inner-working angles.} Inner-working angles as low as $1\lambda/D$ can be reached using current designs.
\item \emph{High throughput.} These designs reach a maximum throughput of 50\% and 30\% for a clear and the VLT aperture, respectively.
\item \emph{High contrast.} Starlight can be suppressed to $<3\times10^{-5}$ for the 6 fibers surrounding the star over the full spectral bandwidth until the throughput starts to drop.
\item \emph{Broad spectral bandwidth.} This suppression is achieved over the full 20\% spectral bandwidth.
\item \emph{Robust against tip-tilt errors.} The SCAR coronagraph is stable against $\sim 0.1\lambda/D$ rms tip-tilt errors upstream due to the use of second order nulling on the fibers.
\item \emph{Residual speckle suppression.} Residual speckles are are reduced by $\sim3\times$ in intensity, due to the coupling efficiency of a random electric field into the single-mode fibers.
\end{enumerate}
All advantages can be obtained into a single \changed{SCAR} design. All these advantages make this coronagraph a prime candidate for future upgrades of extreme AO systems. In particular, the SCAR coronagraph is perfectly suited for spectral characterization of Proxima b: it satisfies all coronagraphic requirements set by \cite{lovis2017atmospheric}. A companion paper \citep{haffert2017singlemode} provides a tolerancing study for this specific application.

Future research will explore active control of the fiber throughput of the SCAR coronagraph. Application of the SCAR methodology to other coronagraphs is also left for future research. An interesting example in this case is the design of a Lyot-plane mask for a conventional Lyot or vortex coronagraph, akin to~\cite{ruane2015lyot}. Even optimizing the focal-plane mask itself might be realizable for the fiber array in these coronagraphs~\citep{ruane2015nodal}.

Another avenue is the use of photonic technologies to further process the light in the fibers. A simple example is the use of fiber Bragg gratings for filtering the atmospheric OH lines~\citep{trinh2013gnosis}. Another example is building a phase-shifting interferometer of the 6 fibers. This will provide information about the coherence of the light in each of the fibers w.r.t. the star, and would allow for synchronous interferometric speckle subtraction~\citep{guyon2004imaging}.

\begin{acknowledgements}
Por acknowledges funding from The Netherlands Organisation for Scientific Research (NWO) and the S\~{a}o Paulo Research Foundation (FAPESP). Haffert acknowledges funding from research program VICI 639.043.107, which is financed by NWO.
\end{acknowledgements}

\bibliographystyle{aa}
\bibliography{references.bib}

\end{document}